\documentclass[journal,twoside,web]{ieeecolor}
\usepackage{tmi}
\usepackage{cite}
\usepackage{amsmath,amssymb,amsfonts}
\usepackage{algorithmic}
\usepackage{graphicx}
\usepackage{textcomp}
\usepackage{mathtools}
\usepackage{algorithm}
\usepackage{textcmds}
\usepackage{url}

\definecolor{revisedColor}{rgb}{0,0,1}

\def\BibTeX{{\rm B\kern-.05em{\sc i\kern-.025em b}\kern-.08em
    T\kern-.1667em\lower.7ex\hbox{E}\kern-.125emX}}
\expandafter\def\expandafter\normalsize\expandafter{%
    \normalsize%
    \setlength\abovedisplayskip{5pt plus 2pt minus 2pt}%
    \setlength\belowdisplayskip{5pt plus 2pt minus 2pt}%
    \setlength\abovedisplayshortskip{3pt plus 1pt minus 1pt}%
    \setlength\belowdisplayshortskip{3pt plus 1pt minus 1pt}%
}

\begin{document}
\title{A General-Purpose Adaptable Foundation Model for Total-Body PET Kinetic Modelling}
\author{Yun Zhao, Qinlin Gu, Andrew J. Reader, Yanan Fan, Georgios I. Angelis, Katya Amadita, Kevin J. Chung, Xiaoyu Duan, Benjamin A. Spencer, Guobao Wang, Steven R. Meikle
\thanks{This work is supported by the Australian Research Council (Discovery Projects grant DP230102070) and the US National Institutes of Health (Research Project grant R01EB033435).}
\thanks{Yun Zhao (yun.zhao@sydney.edu.au) and Katya Amadita are with the School of Health Sciences and the Brain and Mind Centre, the University of Sydney, Australia.
Qinlin Gu is with the School of Mathematics and Statistics and the Brain and Mind Centre, the University of Sydney, Australia.
Andrew J. Reader is with the School of Biomedical Engineering and Imaging Sciences, King’s College London, UK.
Yanan Fan is with CSIRO Technology, Australia.
Georgios I. Angelis is with the School of Biomedical Engineering and Sydney Imaging, the University of Sydney, Australia.
Kevin J. Chung is with the Athinoula A. Martinos Center, Department of Radiology, Massachusetts General Hospital and Harvard Medical School, USA.
Xiaoyu Duan, Benjamin A. Spencer, and Guobao Wang are with the Department of Radiology, University of California Davis Health, USA.
Steven R. Meikle is with the School of Health Sciences, Sydney Imaging and the Brain and Mind Centre, the University of Sydney, Australia.}
}

\maketitle

\begin{abstract}
Voxel-level kinetic modeling in total-body dynamic positron emission tomography (TB-PET) enables precise physiological quantification but complicates parameter uncertainty estimation. While generative consistency models (CMs) can rapidly estimate Bayesian posterior distributions, they are constrained by type of tracers, kinetic models, and protocols. Training them from scratch whenever a new tracer, model or protocol is used can be computationally intensive. To address this, we introduce LoRACM, a parameter-efficient framework utilizing low-rank adaptation (LoRA) to transfer a pretrained foundation CM to novel TB-PET scenarios.
We evaluated three adaptation scenarios using simulated and human data: scanner and protocol transfer for FDG two-tissue compartment modeling; FP-CIT simplified reference tissue modeling; and high-temporal-resolution FDG blood flow imaging via the adiabatic approximation to tissue homogeneity model. Each adaptation was trained in less than 3 hours on a single NVIDIA A100 GPU using 50,000 simulated pairs. In simulations, LoRACM showed strong agreement with results from Markov Chain Monte Carlo (MCMC, which we use as reference posterior distributions), yielding median Kullback–Leibler divergences of 0.05–0.11 and median absolute percentage errors below 1.2\% for each kinetic parameter of interest. In human data, LoRACM generated total-body $K_i$ maps comparable to the ones made by non-negative nonlinear least squares (NNLS) fitting and a consistency model trained from scratch. Striatal dopamine binding potential estimates predicted by LoRACM correlated with NNLS estimates ($r = 0.968$–$0.988$). LoRACM produced total-body blood flow imaging with FDG matching the one made by NNLS. 
These underscore the potential of adapting a pretrained CM with LoRA for fast and reliable  posterior parameter inference for kinetic models across distinct TB-PET applications.
\end{abstract}

\begin{IEEEkeywords}
Generative AI, Diffusion models, Transfer learning, Kinetic modelling, Positron emission tomography
\end{IEEEkeywords}
\vspace{-18pt}
\section{Introduction}
Dynamic positron emission tomography (PET) enables physiological quantification through kinetic modelling of radiotracer delivery, exchange, and retention, supporting measurements such as glucose metabolism and receptor binding beyond static uptake \cite{phelps1979tomographic,lammertsma1996simplified,dias2022normal,wang2022total}.
Total-body and long-axial-field-of-view systems provide increased sensitivity and simultaneous multi-organ coverage \cite{badawi2019first,pantel2020pennpet,prenosil2022performance}, enabling concurrent acquisition of tissue time activity curves (TACs) and image-derived input functions \cite{zhang2020total}.
Translating this acquisition capability into physiological measurements requires parameter estimation methods that accommodate both large image volumes and diverse kinetic behaviour.

Established approaches for PET kinetic parameter estimation in dynamic and total-body PET include standard compartmental modelling typically fitted via non-negative nonlinear least squares (NNLS) \cite{phelps1979tomographic,wang2022total}, alongside graphical and linearized methods such as Patlak and Logan analyses \cite{patlak1983graphical,logan1990graphical}, generalized Patlak modelling \cite{karakatsanis2015generalized}, spectral analysis, and basis-function reference-tissue methods \cite{cunningham1993spectral,gunn1997parametric}. Direct parametric reconstruction further integrates kinetic modelling with the PET measurement model to improve statistical efficiency \cite{wang2009generalized,zhang2020total}. However, these approaches primarily provide point estimates of kinetics rather than full kinetic parameter posterior distributions \cite{gelman1995bayesian}.
While Bayesian inference enables full characterization of posterior distributions, sampling via Markov chain Monte Carlo (MCMC) demands repeated model evaluations and convergence assessments for each TAC \cite{brooks2011handbook}, imposing a substantial computational burden on total-body imaging due to the large number of voxels. Likelihood-free alternatives such as vPET-ABC require careful calibration of the accepting threshold \cite{gu2026vpetabc}.

Deep learning methods accelerate PET kinetic modelling via supervised \cite{zaker2022direct} and self-supervised approaches, such as those by De Benetti et al.\ \cite{de2023self} and Tehlan and Wendler \cite{tehlan2025physiological}. However, these methods primarily yield parameter point estimates without uncertainty quantification.  
Generative methods were introduced to facilitate posterior estimation in dynamic PET. Liu et al.\ demonstrated conditional variational autoencoders for kinetic model parameter estimations \cite{liu2023posterior}. Following developments in denoising diffusion probabilistic models \cite{ho2020denoising,nichol2021improved}, Djebra et al.\ applied an improved diffusion model to brain PET for kinetic posterior estimation, reporting more than 230-fold faster computation than MCMC \cite{djebra2025bayesian}.
Recently, consistency models \cite{song2023consistency} have emerged and adapted for fast simulation-based inference \cite{schmitt2024consistency}. Our previous work introduced a consistency model (CM) for voxel-wise PET kinetic posterior estimation and total-body parametric imaging \cite{zhao2026generative}, achieving MCMC-level accuracy while being over five orders of magnitude faster.

However, diffusion models are locked to specific protocols, tracers, and kinetic models. Training them from scratch for new scenarios is computationally intensive, especially in clinical settings. Parameter-efficient transfer learning offers a solution to this bottleneck \cite{houlsby2019parameter}. Low-rank adaptation (LoRA) retains pretrained weights while learning updates through low-rank matrices \cite{hu2021lora}. This study introduces LoRA for CM (LoRACM) by adapting the pretrained CM to different scenarios using LoRA.
We evaluated LoRACM across three adaptation tasks: 1) a new protocol for $^{18}\mathrm{F}$-fluorodeoxyglucose (FDG) two-tissue compartment modeling (2TCM)~\cite{phelps1979tomographic}, 2) $^{18}\mathrm{F}$-FP-CIT simplified reference tissue modeling (SRTM)~\cite{lammertsma1996simplified}, 3) and high-temporal-resolution FDG with the adiabatic approximation to tissue homogeneity (AATH) modeling~\cite{lawrence1998adiabatic}. 
With minimal fine-tuning overhead (3~h on a NVIDIA A100 GPU), LoRACM recovered reference MCMC posteriors in simulated dataset and generated clinical parametric images consistent with conventional NNLS across the tasks. These results demonstrate that LoRACM can transfer pretrained consistency models across distinct PET kinetic modeling applications, lowering the barrier to deploying kinetic posterior estimation for new tracers and protocols.
LoRACM is integrated into our open-source PET-TRACER toolbox (\url{https://github.com/yundumbledore/PET-TRACER}), featuring a graphical user interface to support accessible deployment across the PET community.

\section{Methods}\label{methods}
\subsection{Pretrained Consistency Model for Kinetic Inference}
\label{sec:pretrained_cm}

\begin{figure*}[t!]
    \centering
    \includegraphics[width=0.75\textwidth]{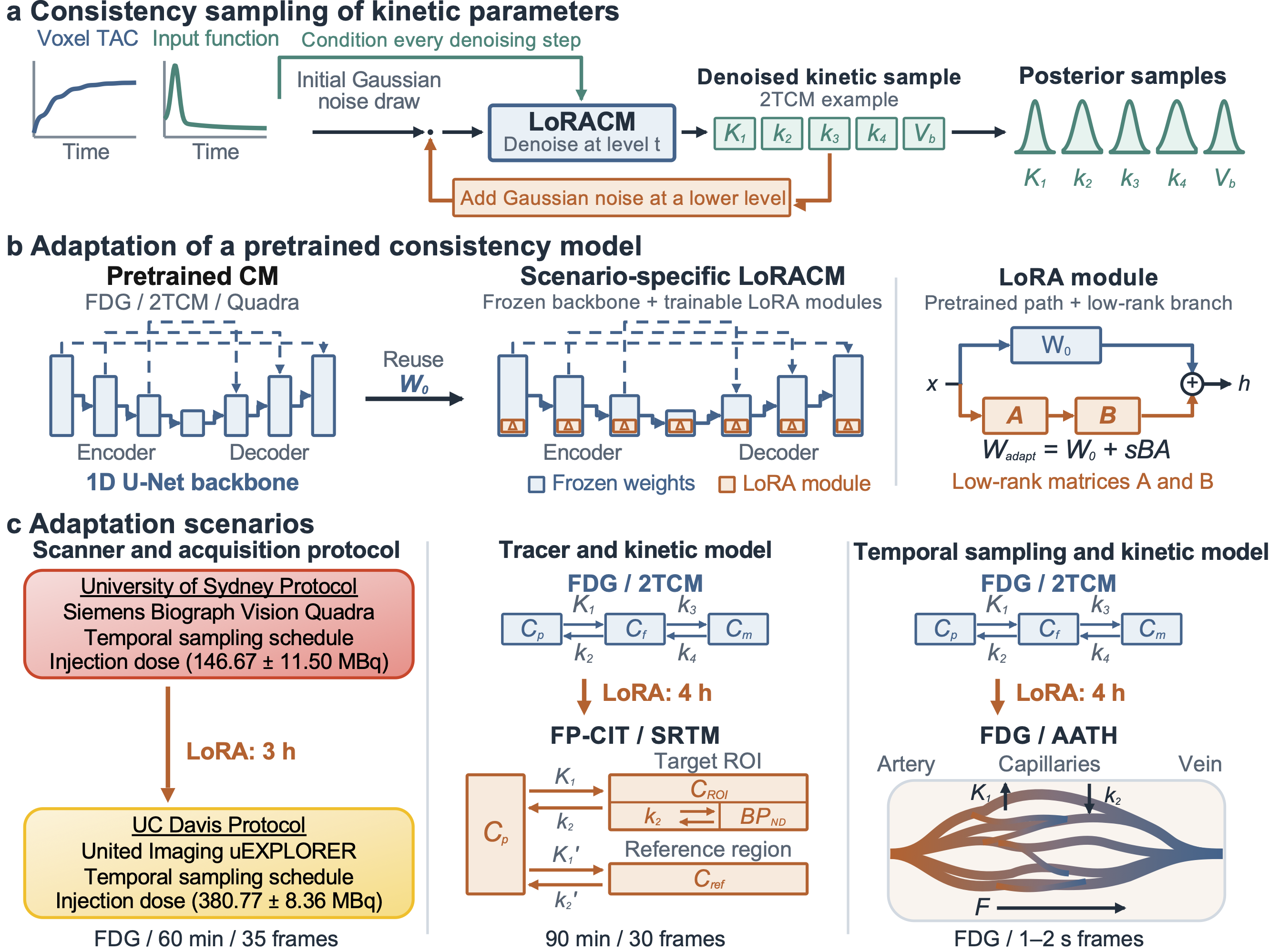}
    \caption{Overview of LoRACM framework. (a) Conditional consistency sampling of kinetic parameters from voxel TACs and input functions. (b) Low-rank adaptation of the pretrained 1D U-Net. (c) Schematic of adaptation scenarios.
    }
    \label{fig: Framework}
\end{figure*}

We adopt the CM developed in our previous work\cite{zhao2026generative}, retaining its formulation and backbone architecture. 
This CM was trained for $^{18}\mathrm{F}$-FDG kinetics with 1,000,000 simulated training pairs ($\mathbf{k}, \mathbf{y}$) in 3 days and can predict the posterior samples of $p(\mathbf{k}|\mathbf{y})$ where $\mathbf{k}=(K_1,k_2,k_3,k_4,V_b)^{\mathsf{T}}$ and $\mathbf{y}$ incorporates the voxel-level TAC and corresponding arterial input function (AIF).
The CM employs a one-dimensional U-Net to map a noisy kinetic-parameter vector to a denoised estimate, conditioned on the measured curves and noise level. During model training, predictions at adjacent noise levels are encouraged
to agree using an exponential-moving-average target network. The conditioning information is incorporated throughout the U-Net.
As illustrated in Fig.\ref{fig: Framework}(a), posterior sampling begins with a Gaussian noise draw and alternates denoising with noise injection at progressively lower noise levels. Repeating this procedure with independent noise draws yields samples from the learned conditional posterior. Full architectural, training, and sampling details are provided in\cite{zhao2026generative}. The pretrained CM provides the starting point for the LoRA described next.

\subsection{Low-Rank Adaptation of the Consistency Model}
\label{sec:lora}

We adapt the pretrained CM to each application scenario using LoRA~\cite{hu2021lora}, yielding a scenario-specific model termed LoRACM. As illustrated in Fig.\ref{fig: Framework}(b), LoRA represents the update to a pretrained weight matrix
$\mathbf{W}_0\in\mathbb{R}^{d_{\mathrm{out}}\times d_{\mathrm{in}}}$ as 
\begin{equation}
\mathbf{W}
=
\mathbf{W}_0+\frac{\alpha}{r}\mathbf{B}\mathbf{A},
\label{eq:lora}
\end{equation}
where
$\mathbf{A}\in\mathbb{R}^{r\times d_{\mathrm{in}}}$ and $\mathbf{B}\in\mathbb{R}^{d_{\mathrm{out}}\times r}$ are trainable matrices, $r$ specifies the adaptation rank, and $\alpha$ controls the update magnitude. The original weight matrix remains frozen, while the adapter introduces $r(d_{\mathrm{in}}+d_{\mathrm{out}})$ trainable parameters.

Adapters (with 0.05 input dropout) are applied to linear, 1D convolutional, and transposed convolutional layers, reshaping the low-rank product to match convolutional kernel dimensions. Matrices $\mathbf{A}$ and $\mathbf{B}$ use Kaiming-uniform~\cite{he2015delving} and zero initialization, respectively, ensuring zero initial change to the network output. To accommodate differing temporal sampling and kinetic model dimensions, the target network is constructed with the required conditioning and parameter vector dimensions: shape-compatible pretrained weights are transferred and frozen, whereas normalization parameters and newly initialized shape-incompatible layers are trained alongside the adapters. Fine-tuning retains our consistency training objective~\cite{song2023consistency,zhao2026generative}, where identical Gaussian noise is added to the standardized kinetic-parameter vector at two adjacent noise levels, and the trainable student network is optimized by minimizing the mean squared error between its higher-noise prediction and the stop-gradient prediction of an exponential-moving-average target network at the lower noise level.

For each adaptation scenario, the adaptation rank and training-set size were selected through a grid search. Candidate models were evaluated on simulated data by comparing their parameter posterior distributions with those obtained using MCMC. Hyperparameter selection was based on
$D_{\mathrm{KL}}(p_{\mathrm{MCMC}}\Vert
p_{\mathrm{LoRACM}})$.
The search considered $r\in\{2,4,8,16,32,64\}$, and training subset size ranging from 1\%--25\% of the original pretraining dataset. 
The selected settings were $(r,\alpha)=(4,8)$ for adaptation scenario~1, and $(r,\alpha)=(8,16)$ for adaptation scenario~2 and 3.

\subsection{Adaptation Scenarios and Kinetic Models}\label{sec:kinetic_models}
Three adaptation scenarios are illustrated in Fig.\ref{fig: Framework}(c) and they evaluate adaptations across scanners, acquisition protocols, tracers, and kinetic models.
All scenarios start from the same pretrained CM that we developed in \cite{zhao2026generative}. The target kinetic models and inferred parameters are described below.

\subsubsection{Scenario 1: Scanner and Acquisition-Protocol Adaptation}
For total-body parametric imaging, we adapted the CM from the University of Sydney protocol—on which it was pretrained—to the UC Davis protocol, evaluating CM transfer across distinct temporal sampling schedules and cohorts while maintaining FDG tracer and the reversible 2TCM. The model describes tracer exchange between plasma, free-tissue, and metabolized compartments, accounting for blood arrival delay $t_d$
\begin{equation}
\begin{aligned}
\frac{dC_f}{dt}
    &= K_1 C_p(t - t_d)-(k_2+k_3)C_f+k_4 C_m,\\
\frac{dC_m}{dt}
    &= k_3 C_f-k_4 C_m,\\
C_T(t)
    &= (1-V_b)\bigl[C_f(t)+C_m(t)\bigr]+V_b C_b(t-t_d),
\end{aligned}
\end{equation}
where $C_p$ and $C_b$ denote plasma and whole-blood activity concentrations. The inferred parameters are $(K_1,k_2,k_3,k_4,V_b,t_d)^{\mathsf T}$, representing delivery, efflux, phosphorylation, dephosphorylation, blood volume fraction, and arrival delay, respectively \cite{phelps1979tomographic}. 
From the posterior samples of these micro-parameters, the net influx rate is derived as $K_i = K_1 k_3 / (k_2 + k_3)$.
The same input curve is used for $C_p$ and $C_b$ in our implementation.

\subsubsection{Scenario 2: Tracer and Reference-Tissue Model Adaptation}
To evaluate adaptation to non-invasive kinetic modeling, the pretrained CM was adapted to SRTM~\cite{lammertsma1996simplified} using the dopamine transporter (DAT) ligand $^{18}\mathrm{F}$-FP-CIT. Reference-tissue methods avoid invasive arterial blood sampling by using a reference region devoid of specific binding. Here, the cerebellum TAC serves as an indirect input function. Under the assumption that target and reference tissues exhibit single compartment kinetics with equal non-displaceable distribution volumes, SRTM describes the target tissue TAC as:
\begin{equation}
\begin{aligned}
C_T(t)
    &= R_1 C_R(t)
       +(k_2-R_1 k_{2a})
       \bigl[C_R\otimes e^{-k_{2a}t}\bigr](t),\\
k_{2a}
    &= \frac{k_2}{1+BP_{\mathrm{ND}}},
\end{aligned}
\end{equation}
where $\otimes$ denotes temporal convolution and $C_R$ is the reference-tissue TAC \cite{lammertsma1996simplified}. The inferred parameters are $(R_1,k_2,BP_{\mathrm{ND}})^{\mathsf T}$. $R_1=K_1/K_1'$ denotes delivery relative to the reference tissue, $k_2$ denotes target-tissue efflux, and $BP_{\mathrm{ND}}$ is the binding potential. Here, the cerebellar TAC was used as the reference input.

\subsubsection{Scenario 3: Adaptation to High-Temporal-Resolution Blood-Flow Imaging}
The pretrained CM was adapted to AATH modelling to early, high-temporal-resolution FDG data \cite{chung2025quantitative}. This scenario changes both the temporal sampling schedule and kinetic model to estimate blood flow.
The AATH model accounts for a mean transit time from artery to vein. Its impulse response function is defined as
\begin{equation}
\begin{aligned}
R(t)
    &=
    \begin{cases}
    0, & t<0,\\
    F, & 0\leq t<T_c,\\
    K_1 e^{-k_2(t-T_c)}, & t\geq T_c,
    \end{cases}\\
C_T(t)
    &= \bigl[C_a(t-t_d)\otimes R\bigr](t).
\end{aligned}
\end{equation}
The inferred parameters are $(K_1,k_2,F,T_c,t_d)^{\mathsf T}$, where $F$ is blood flow, $T_c$ is mean vascular transit time, and $t_d$ is arterial arrival delay. Blood-to-tissue transport satisfies $K_1=FE$, with $E$ denoting extraction fraction \cite{chung2025quantitative}.

\subsection{Training and Evaluation Procedures}
\label{sec:training_evaluation}

Scenario-specific simulation datasets were generated using the corresponding kinetic forward models. Kinetic parameters were sampled from uniform priors over ranges based on published studies~\cite{zhao2026generative,yaqub2007quantification,
chung2025quantitative}.
The numerical bounds are specified in Table \ref{tab:parameter_bounds}. Input functions were derived from patient measurements: image-derived input functions from ascending aorta for Scenarios~1 and~3 and cerebellar reference-tissue TACs for Scenario~2.
\begin{table}[t]
\centering
\caption{Uniform parameter-sampling bounds for the
three adaptation scenarios.}
\label{tab:parameter_bounds}
\footnotesize
\setlength{\tabcolsep}{4pt}
\renewcommand{\arraystretch}{1.05}
\begin{tabular*}{\columnwidth}
{@{\extracolsep{\fill}}llcc@{}}
\hline
Parameter & Unit & Lower & Upper \\
\hline
\multicolumn{4}{l}{\textit{Scenario 1: FDG--2TCM}} \\
$K_1$ & mL\,min$^{-1}$\,cm$^{-3}$ & 0.0001 & 2.00 \\
$k_2$ & min$^{-1}$                & 0.0001 & 2.00 \\
$k_3$ & min$^{-1}$                & 0.0001 & 0.50 \\
$k_4$ & min$^{-1}$                & 0.0001 & 0.20 \\
$V_b$ & Dimensionless            & 0.001  & 1.00 \\
\hline
\multicolumn{4}{l}{\textit{Scenario 2: FP-CIT--SRTM}} \\
$R_1$     & Dimensionless & 0.0001    & 4.00 \\
$k_2$     & min$^{-1}$    & 0.0001  & 0.50 \\
$BP_{ND}$ & Dimensionless & 0.0001 & 7.00 \\
\hline
\multicolumn{4}{l}{\textit{Scenario 3: FDG--AATH}} \\
$E  $ &        & 0.0001  & 1.00 \\
$k_2$ & min$^{-1}$                 & 0.0001     & 2.50 \\
$F$   & mL\,min$^{-1}$\,cm$^{-3}$  & 0.0001  & 3.50 \\
$T_c$ & min                       & 0.016 & 0.85 \\
$t_d$ & min                       & 0     & 0.35 \\
\hline
\end{tabular*}
\end{table}
To approximate the non-negative, positively skewed measurement variability characteristic of iterative Expectation-Maximization (EM) reconstruction in dynamic PET, simulated TACs were corrupted using the log-normal noise model \cite{barrett1994noise,wilson1994noise} given by,
\begin{equation}
\widetilde{C}_T(t_i)
=
\bar{C}_T(t_i) \exp\!\left( \eta_i \epsilon_i - \frac{1}{2}\eta_i^2 \right),
\qquad
\epsilon_i\overset{\mathrm{i.i.d.}}{\sim}\mathcal{N}(0,1),
\label{eq:measurement_noise}
\end{equation}
where $\bar{C}_T(t_i) = \max\{C_T(t_i), \varepsilon\}$, and the log-space standard deviation $\eta_i$ is related to the frame standard deviation $\sigma_i$ by
\begin{equation}
\eta_i = \sqrt{\ln\!\left( 1 + \frac{\sigma_i^2}{\bar{C}_T(t_i)^2} \right)},
\qquad
\sigma_i
=
\ell
\sqrt{
\frac{\bar{C}_T(t_i) e^{-\lambda t_i}}{\Delta t_i}
}
e^{\lambda t_i},
\label{eq:noise_sd}
\end{equation}
with decay constant $\lambda = \ln 2 / T_{1/2}$. Here, $\ell$ controls the noise level ($3 < \ell < 7$), $T_{1/2} = 109.8$~min is the half-life of $^{18}$F, and $\varepsilon = 10^{-8}$ ensures numerical stability and strict positivity for the log-normal transformation.
For each scenario, we generated 100,000 training pairs, 20,000 validation pairs and 100 simulated test samples using the same prior distribution as the training data.

All scenarios used AdamW optimizer with an initial learning rate of $10^{-4}$ and a batch size of 512. The learning rate was reduced when validation loss plateaued, and the checkpoint with the lowest validation loss was selected. Adaptation was performed on a NVIDIA A100 GPU and required less than 3~h for every scenario.

MCMC was used to provide reference posterior samples using the No-U-Turn Sampler in NumPyro\cite{hoffman2014no}. The sampling procedure used 4 chains with 2,000 warm-up iterations and 2,000 draws. Sampling was repeated with progressively doubled draw counts, up to 16,000, when the monitored $\widehat{R}$ statistic did not fall below 1.1.
NNLS fitting provided a point estimation baseline, and the implementation in our study followed \cite{wang2018dynamic}.

Parameter-wise posterior agreement was evaluated via $D_{\mathrm{KL}}(p_{\mathrm{MCMC}}\Vert p_{\mathrm{LoRACM}})$, and point-estimate accuracy (using posterior medians) for nonzero ground-truth $k_j^{\mathrm{true}}$ via absolute percentage error, $\mathrm{APE}_j = 100\,|\operatorname{median}(k_j\mid\mathbf{y}) - k_j^{\mathrm{true}}| / |k_j^{\mathrm{true}}|$.

\subsection{Clinical Datasets and Acquisition Protocols}\label{dataset}
\subsubsection{Total-Body FDG PET: Scenarios 1 and 3}
Dynamic $^{18}\text{F}$-FDG PET/CT was acquired on a uEXPLORER total-body system from a cohort of 10 subjects scanned twice under two distinct protocols. Low-dose CT (140~kVp, 5--6~mAs) was used for attenuation correction. All PET data were reconstructed using TOF-OSEM (4 iterations, 20 subsets, standard corrections, without PSF) into a $150\times150\times486$ matrix ($4\times4\times4$~$\text{mm}^3$ voxels). For Scenario~1, 60-min dynamic scans from the low-dose session ($183\pm17$~MBq) were reconstructed into 35 frames ($12\times5$~s, $4\times15$~s, $2\times30$~s, $3\times60$~s, $8\times180$~s, $6\times300$~s). For Scenario~3, high-temporal-resolution data from the high-dose session ($304\pm52$~MBq, two subjects) were reconstructed across the first 2~min into 90 frames ($60\times1$~s and $30\times2$~s)~\cite{chung2025quantitative}. Image-derived arterial input functions were obtained from the ascending aorta.

\subsubsection{Longitudinal FP-CIT PET: Scenario 2}
Dynamic 90-min FP-CIT PET (148$\pm$12.5~MBq) was acquired on a Siemens Biograph Vision Quadra from two mild Parkinson's disease patients at baseline, 6, and 12 months during an exercise intervention. Low-dose CT (Sn140~kVp, 6~mAs) was used for attenuation correction. Brain images (440$\times$440 matrix, zoom 2) were extracted from whole-body data following PSF-TOF OP-OSEM reconstruction (8 iterations, 5 subsets, standard corrections, no post-filtering) into 30 dynamic frames ($1\times15$~s, $3\times5$~s, $3\times10$~s, $2\times30$~s, $3\times60$~s, $2\times150$~s, $16\times300$~s). Using PMOD, data underwent frame-by-frame motion correction and probabilistic atlas registration. The dynamic series was denoised using spatiotemporal HYPR-LR filtering\cite{christian2010dynamic}.
TACs were extracted from the bilateral caudate and putamen using a cerebellum reference.

\section{Results}
\subsection{Adaptation to UC Davis Protocol}
\begin{figure}[!t]
    \centering
    \includegraphics[width=\columnwidth]{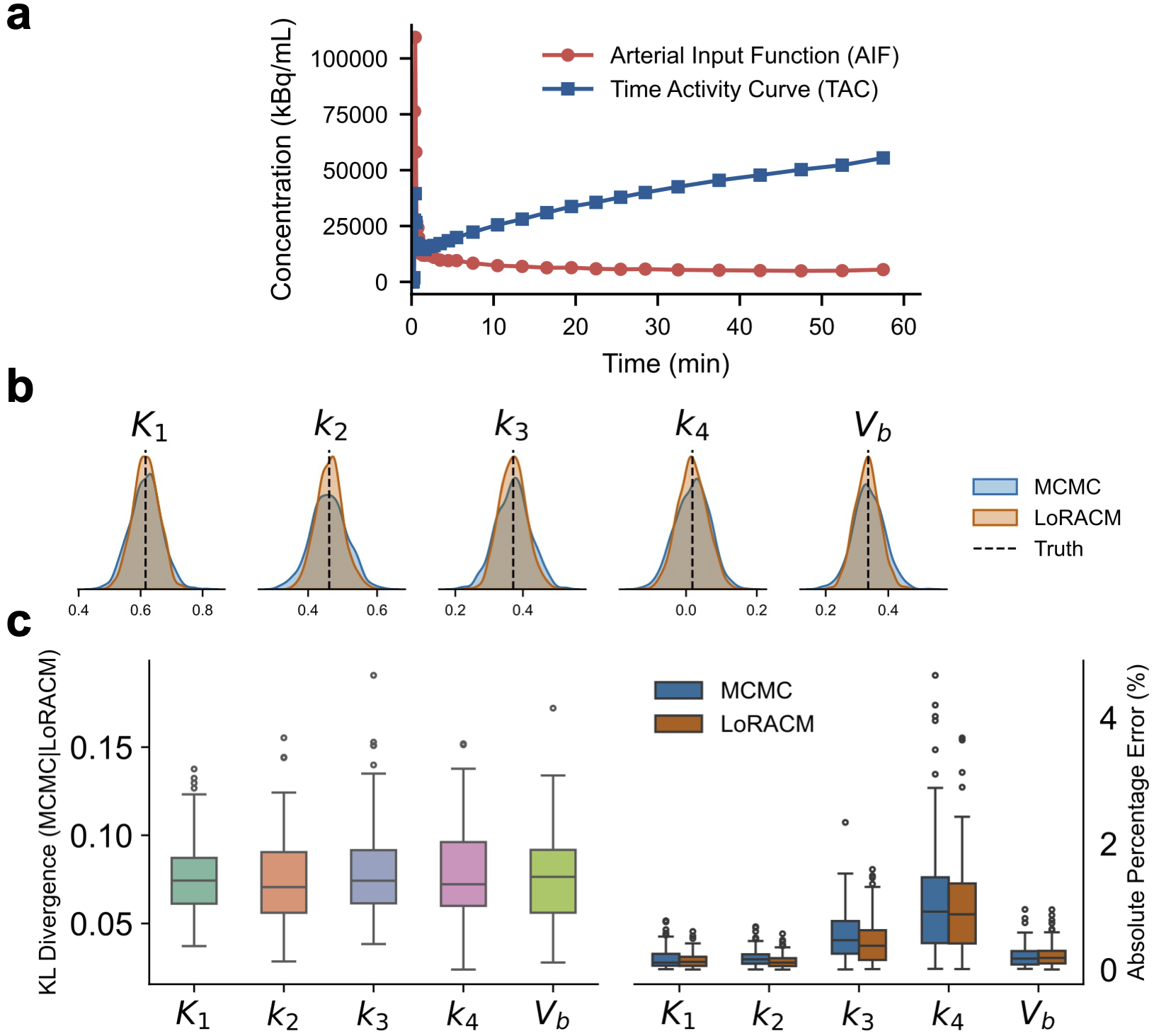}
    \caption{Simulation evaluation for Scenario 1. (a) An example simulated AIF and tissue TAC. (b) Posteriors predicted by MCMC and LoRACM with dashed lines indicating ground truth. (c) Parameter-wise posterior KL divergence (left) and absolute percentage error of posterior medians (right) across 100 simulated test cases.}
    \label{fig: C1_simulation_study}
\end{figure}

Fig.~\ref{fig: C1_simulation_study} evaluates LoRACM on simulated data following adaptation to the UC Davis protocol. For a representative simulated TAC and AIF in Fig.~\ref{fig: C1_simulation_study}(a), the posterior distributions predicted by LoRACM closely match the MCMC reference across all five kinetic parameters, with the ground-truth values centered near the posterior modes (Fig.~\ref{fig: C1_simulation_study}(b)). Across the 100 simulated test cases, each parameter yielded a median KL divergence of 0.07--0.08 (Fig.~\ref{fig: C1_simulation_study}(c), left). 
Furthermore, the APE distributions of the posterior medians were statistically indistinguishable between LoRACM and MCMC (Fig.~\ref{fig: C1_simulation_study}(c), right; paired $t$-test, $t = 0.25$, $p = 0.80$). Both methods achieved median APEs below 1\% across all parameters, with $K_1$, $k_2$, and $V_b$ exhibiting the highest precision and $k_4$ showing the largest error and dispersion.

\begin{figure}[!t]
    \centering
    \includegraphics[width=0.975\columnwidth]{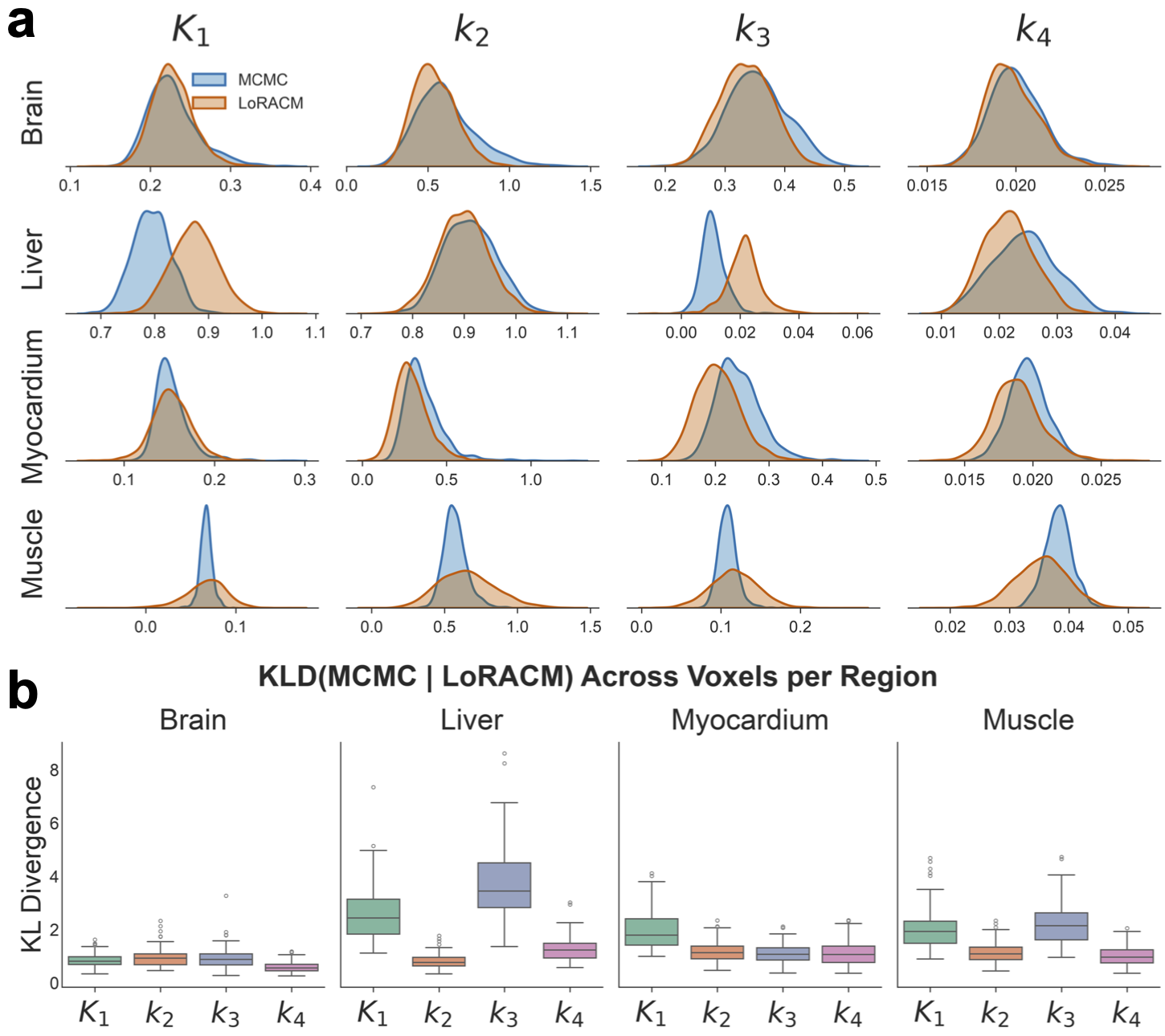}
    \caption{Posterior evaluation on real FDG data in Scenario 1. (a) Posterior predictions by MCMC and LoRACM for a voxel in the brain, liver, myocardium, and muscle, respectively. (b) KL divergence between posterior predictions by MCMC and LoRACM across voxels within each region.}
    \label{fig: C1_roi_study}
\end{figure}

Evaluation on real human FDG data revealed tissue dependent posterior agreement. In Fig.\ref{fig: C1_roi_study}(a), LoRACM reproduced the principal locations and shapes of the MCMC posteriors for the brain and myocardium voxels. Larger differences were observed in the liver, particularly for $K_1$ and $k_3$, where the LoRACM distributions shifted toward higher values. In muscle, the posterior locations were broadly consistent, but LoRACM produced wider distributions, most visibly for $k_2$ and $k_4$.
The voxel-wise KL divergence distributions in Fig.\ref{fig: C1_roi_study}(b) showed the lowest and most consistent divergence in the brain, while liver $K_1$ and $k_3$ exhibited larger discrepancies. Myocardium and muscle had low divergence for all kinetic parameters with median values below $2.0$.

\begin{figure}[!t]
    \centering
    \includegraphics[width=0.85\columnwidth]{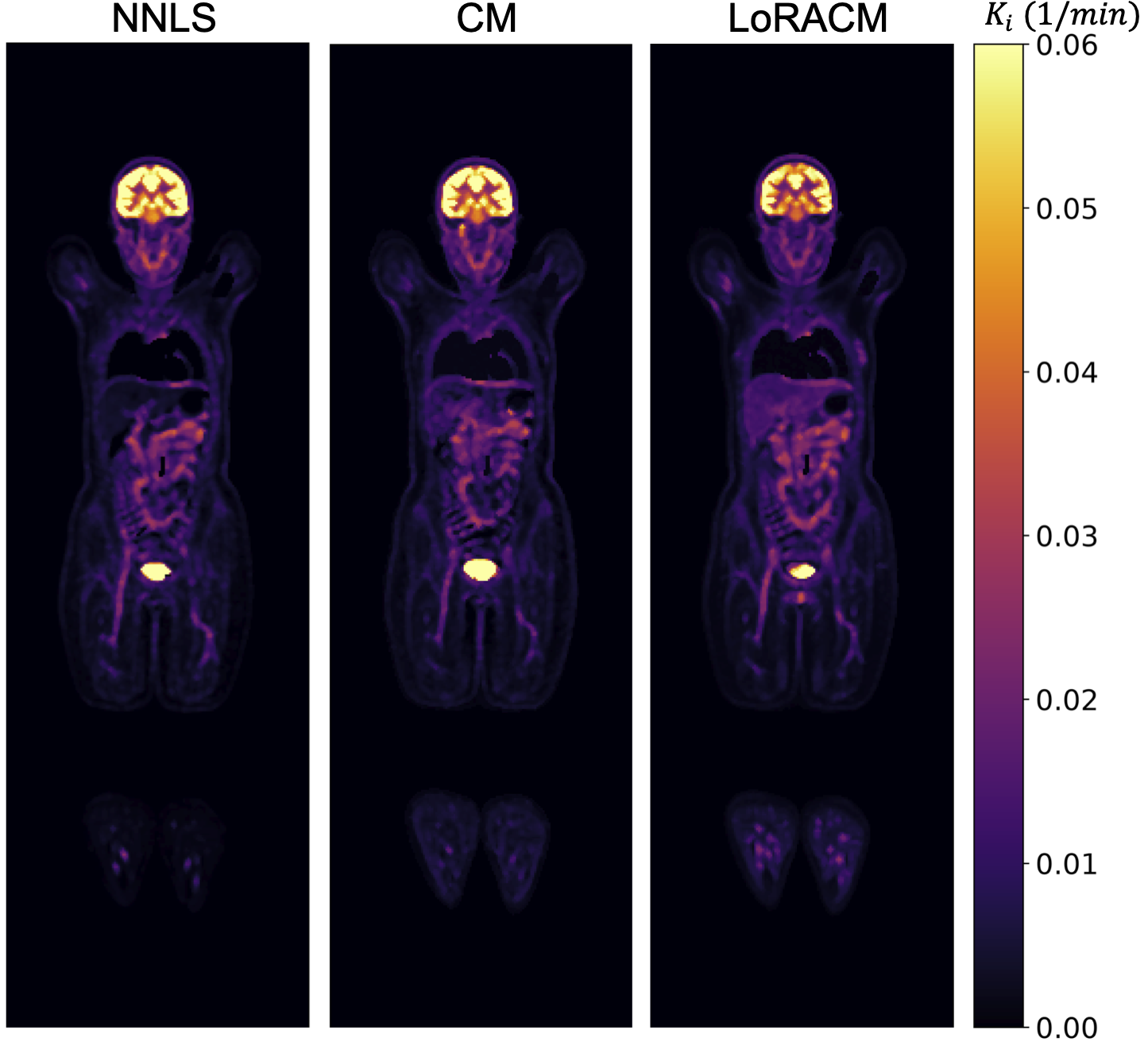}
    \caption{Total-body FDG net influx rate ($K_i$) maps estimated using NNLS (left), a consistency model trained from scratch for this protocol (CM, middle), and the adapted LoRACM (right). The NNLS map displays direct point estimates, whereas the CM and LoRACM maps show the voxel-wise posterior medians.}
    \label{fig: C1_parametric_imaging}
\end{figure}

Fig.~\ref{fig: C1_parametric_imaging} compares total-body $K_i$ maps generated from NNLS point estimates against the voxel-wise posterior medians predicted by the from-scratch CM and LoRACM for this new protocol. 
All three methods reproduced the principal spatial patterns, including elevated cerebral $K_i$, heterogeneous abdominal distributions, and relatively low values in the lungs and peripheral soft tissues. 
Within the abdomen, the two CM-based maps retained the major structures visible with NNLS, although regional intensities and contrast have slight differences, particularly in the liver where both CM methods showed slightly higher $K_i$. Peripheral soft tissue and distal lower limb signals were more apparent in both CM-based maps than in the NNLS map. 

\subsection{Adaptation to Binding Potential Imaging}
Table \ref{tab:srtm_aath_evaluation} Scenario 2 summarizes posterior agreement and point estimation accuracy after adaptation to SRTM with FP-CIT. Median KL divergences between LoRACM and MCMC were 0.07, 0.09, and 0.05 for $R_1$, $k_2$, and $BP_{\mathrm{ND}}$, respectively, with the closest posterior agreement observed for $BP_{\mathrm{ND}}$. LoRACM achieved median APEs of 0.32\%, 0.70\%, and 0.35\%, compared with 0.30\%, 0.72\%, and 0.36\% for MCMC. This indicates that LoRACM preserved point estimation accuracy relative to the MCMC reference across the three SRTM parameters.



\begin{figure}[!t]
    \centering
    \includegraphics[width=0.7\columnwidth]{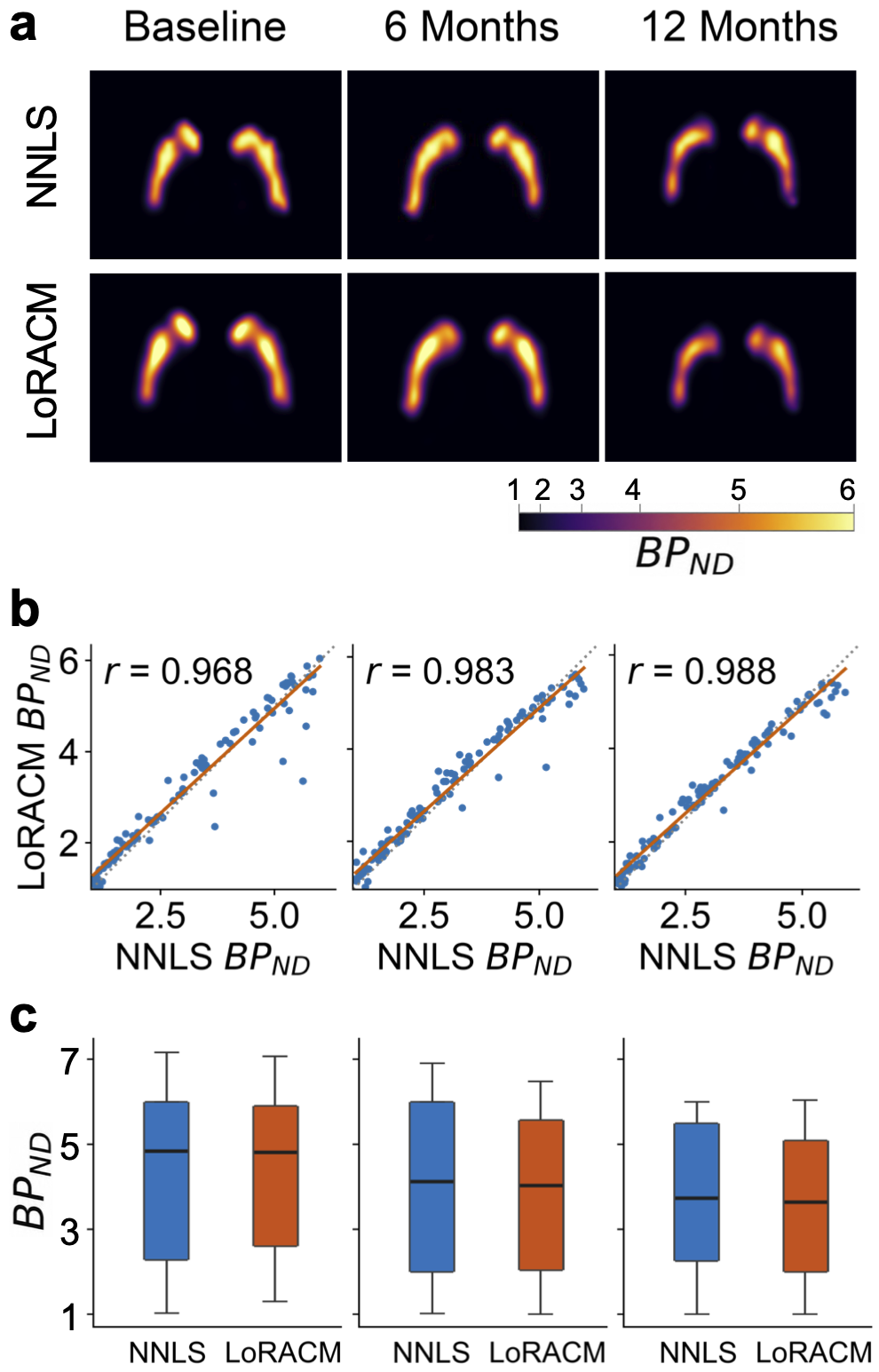}
    \caption{Longitudinal evaluation in Scenario 2. (a) Striatal $BP_{\mathrm{ND}}$ maps from NNLS and LoRACM at baseline, 6 months, and 12 months. (b) Voxel-wise comparisons of estimates from NNLS and LoRACM posterior median, with correlation coefficients $r$, fitted lines (orange), and identity lines (dotted) across the three time points. (c) Corresponding distributions of $BP_{\mathrm{ND}}$ across the three time points.
    }
    \label{fig: C2_real_data_study}
\end{figure}

Fig.\ref{fig: C2_real_data_study} evaluates longitudinal striatal $BP_{\mathrm{ND}}$ estimation on clinical FP-CIT data. The NNLS and LoRACM maps showed similar striatal distributions at baseline and at 6- and 12-month follow-up (Fig.\ref{fig: C2_real_data_study}(a)). Both methods reproduced the progressive reduction in binding potential over time while preserving the spatial pattern of striatal uptake. Voxel-wise estimates were strongly correlated, with $r=0.968$, $0.983$, and $0.988$ at the three time points, respectively
(Fig.\ref{fig: C2_real_data_study}(b)). 
The corresponding distributions of binding potential estimates (Fig.\ref{fig: C2_real_data_study}(c)) showed similar medians and interquartile ranges between methods at each time point, with both capturing the downward shift in $BP_{\mathrm{ND}}$ over time.

\begin{table}[!t]
\centering
\caption{Simulation evaluation for scenarios 2 and 3.
Values are reported as median [first quartile, third quartile].}
\label{tab:srtm_aath_evaluation}
\footnotesize
\setlength{\tabcolsep}{3pt}
\renewcommand{\arraystretch}{1.15}
\begin{tabular*}{\columnwidth}
{@{\extracolsep{\fill}}lccc@{}}
\hline
Parameter & KLD & \multicolumn{2}{c}{APE (\%)} \\
\cline{3-4}
 & MCMC$\Vert$LoRACM & MCMC & LoRACM \\
\hline
\multicolumn{4}{l}{\textit{Scenario 2: SRTM}} \\
$R_1$     & 0.07 [0.05, 0.09] & 0.30 [0.15, 0.55] & 0.32 [0.16, 0.58] \\
$k_2$     & 0.09 [0.06, 0.12] & 0.72 [0.55, 1.01] & 0.70 [0.54, 1.02] \\
$BP_{ND}$ & 0.05 [0.03, 0.08] & 0.36 [0.19, 0.52] & 0.35 [0.22, 0.49] \\
\hline
\multicolumn{4}{l}{\textit{Scenario 3: AATH}} \\
$E$       & 0.08 [0.04, 0.13] & 0.54 [0.32, 0.74] & 0.58 [0.39, 0.71] \\
$k_2$     & 0.11 [0.06, 0.17] & 0.61 [0.33, 0.91] & 0.58 [0.30, 0.88] \\
$F$     & 0.08 [0.05, 0.11] & 0.31 [0.21, 0.40] & 0.30 [0.22, 0.41] \\
$T_c$     & 0.09 [0.05, 0.13] & 1.12 [0.78, 1.49] & 1.19 [0.87, 1.58] \\
$t_d$     & 0.10 [0.05, 0.14] & 1.06 [0.59, 1.38] & 1.12 [0.83, 1.47] \\
\hline
\end{tabular*}

\vspace{3pt}
\begin{minipage}{\columnwidth}
\footnotesize
KLD: $D_{\mathrm{KL}}(p_{\mathrm{MCMC}}\Vert
p_{\mathrm{LoRACM}})$.
APE: absolute percentage error.
\end{minipage}
\end{table}

\subsection{Adaptation to Blood Flow Imaging}
\begin{figure}[!t]
    \centering
    \includegraphics[width=\columnwidth]{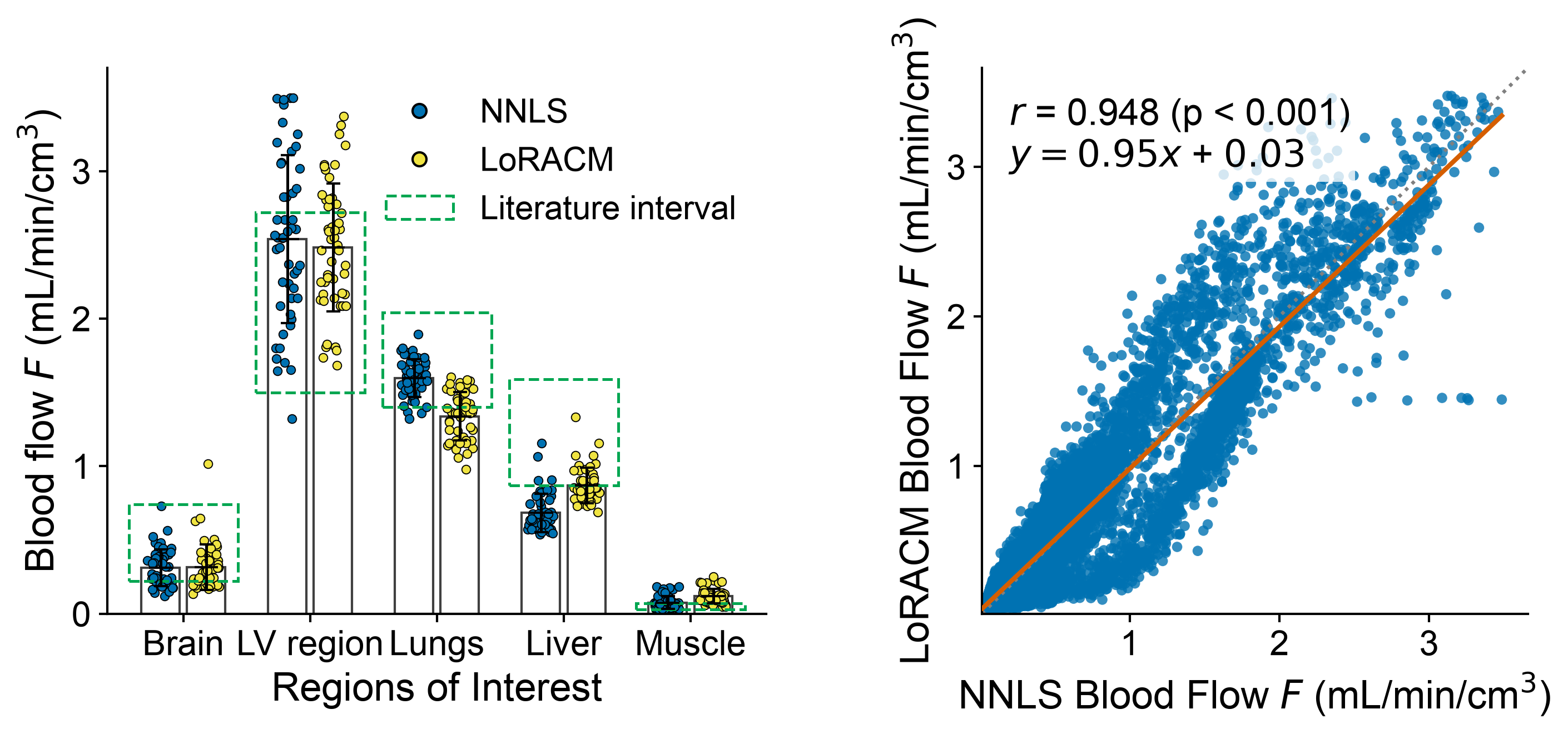}
    \caption{Left: regional estimates of blood flow using NNLS and LoRACM, with published ranges shown in green. Right: voxel-wise comparison of NNLS and LoRACM blood flow estimates, showing the fitted regression (orange) and identity line (dotted).
    }
    \label{fig: C3_roi_study}
\end{figure}

\begin{figure}[!t]
    \centering
    \includegraphics[width=0.85\columnwidth]{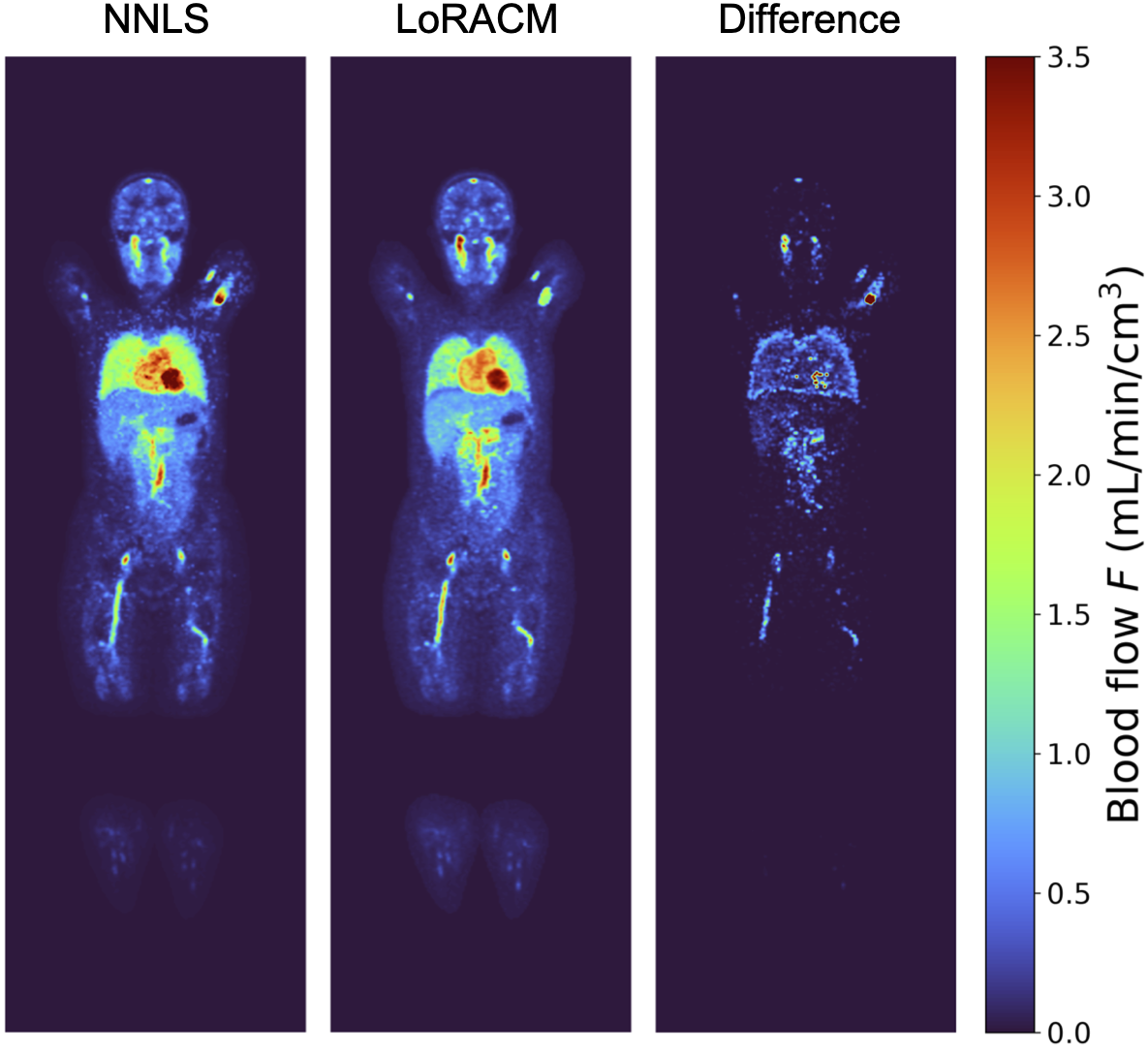}
    \caption{Total-body blood-flow maps from NNLS and LoRACM, with the corresponding voxel-wise squared error map (right).
    }
    \label{fig: C3_parametric_imaging}
\end{figure}

Table \ref{tab:srtm_aath_evaluation} Scenario 3 summarizes the simulation evaluation following adaptation to the AATH model. Median KLD ranged from 0.08 to 0.11 across the five parameters. LoRACM yielded median APEs of 0.58\%, 0.58\%, 0.30\%, 1.19\%, and 1.12\% for $K_1$, $k_2$, $F$, $T_c$, and $t_d$, respectively, closely matching the corresponding MCMC errors. Blood flow exhibited the smallest median APE, whereas vascular transit time and arrival delay showed larger relative errors with both methods.
On real data, LoRACM estimated AATH model parameters from voxel-wise high temporal resolution clinical FDG data. AATH model was fitted to the data with NNLS to infer blood flow as a reference. 
The regional comparisons in Fig.\ref{fig: C3_roi_study} showed similar blood flow distributions between estimates from NNLS and LoRACM, and their estimates were close to published blood flow values in the selected regions \cite{li2023total,chung2025quantitative}. Both methods reproduced the variation from low values in muscle to higher values in the lungs and left ventricular region. The comparison with NNLS estimates showed a strong correlation ($r=0.948$, $p<0.001$), with a regression slope of 0.95.
Fig.\ref{fig: C3_parametric_imaging} compares the total-body blood flow maps estimated by NNLS and LoRACM.
The two maps produced similar spatial patterns, including prominent cardiac and lung signals, visible cerebral perfusion, and lower blood flow throughout much of the peripheral soft tissue.
The LoRACM map retained the principal anatomical features and regional contrast of the NNLS reference.
The discrepancy map shows minor local differences in lungs, cardiac, renal regions, while much of the low-flow background showed trivial deviations.

\section{Discussion}
This study extends our CM framework for kinetic posterior estimation\cite{zhao2026generative} into an adaptable tool for PET parametric imaging. The principal contribution is generalizing generative diffusion models across acquisition protocols, tracers, temporal sampling schedules, kinetic models, and scanners. Requiring only 100,000 simulated training pairs (10\% of the pretraining data), 3~h of fine-tuning, and less than one minute of inference time per coronal slice on a single NVIDIA A100 GPU, LoRACM offers a practical alternative to training CM from scratch. It also provides time efficiency compared to conventional methods like NNLS, which requires at least 8 hours for the same slice and only yields point estimates.

In Scenario 1, LoRACM demonstrated strong agreement with MCMC on the simulated testing dataset. In clinical data, liver $K_1$ and $k_3$ showed distributional shifts between the two methods. This discrepancy stems from model misspecification, as applying a single-input 2TCM to the liver neglects its dual blood supply from the hepatic artery and portal vein \cite{wang2018dynamic,duan2025quantification,wang2021dynamic,zuo2019structural}. Although both methods share the same single-input forward model and prior, they respond differently to this observations. MCMC evaluates the likelihood to identify parameter combinations within the prior that minimize residual error against the observed liver TAC, whereas LoRACM performs inference learned from single-input simulated TACs. Consequently, a clinical dual-input liver TAC lies outside the simulated data manifold, forcing LoRACM to extrapolate and resulting in a shifted posterior relative to MCMC. Because this study primarily focuses on enhancing the generalizability of existing diffusion models across different scenarios, applying distinct kinetic models to specific organs falls outside the scope of this work.
LoRACM was compared with a CM trained from scratch for total-body parametric imaging of $K_i$. The LoRACM $K_i$ map preserved the principal total-body $K_i$ patterns observed in the CM-generated map. This indicates the robust capability of LoRA: requiring less than 3~h of fine-tuning and minimal data, it achieves parametric imaging comparable to a CM trained on 1 million pairs and requiring 3 days.
Scenario 2 evaluated adaptation to SRTM. LoRACM achieved close MCMC agreement for binding potential and demonstrated better stability than NNLS in estimating target-tissue efflux, likely mitigating SRTM's known noise sensitivity\cite{wu2002noise} by learning over bounded parameter distributions. Furthermore, the longitudinal maps successfully captured the observed changes in striatal binding potential over time. This underscores the applicability of LoRACM to clinical research studies in the future.
Scenario 3 adapted to AATH modeling for blood flow estimation. While fitting AATH to $^{11}\text{C}$-Butanol is the standard for approximating blood flow, Chung et al. \cite{chung2025quantitative} demonstrated that early phase, high temporal resolution $^{18}\text{F}$-FDG yields equivalent estimates ($r=0.955$). Our results showed accurate blood flow estimation, whereas vascular transit time ($T_c$) and arrival delay ($t_d$) showed larger errors. Beyond the high noise levels in early short frames, these larger errors reflect the strong collinearity and parameter covariance between $t_d$ and $T_c$ inherent to the AATH formulation \cite{chung2025quantitative}. While Bayesian inference via MCMC or LoRACM does not eliminate this intrinsic parameter covariance, it captures the full joint posterior distribution rather than collapsing onto a single point estimate. On clinical data, NNLS and LoRACM total-body blood flow maps preserved major anatomical features across tissues. Because NNLS serves only as a conventional baseline rather than a ground-truth gold standard, minor local discrepancies between the maps can come from unmodeled dispersion effects \cite{wang2023high} or NNLS converging to arbitrary local optima along the covariant $(t_d, T_c)$ likelihood ridge, whereas LoRACM stabilizes estimation through posterior averaging.

The ability to accommodate different kinetic models is important to total-body PET, where a single model cannot adequately describe all tissues. Wang et al. demonstrated that voxel-wise model selection influence parametric quantification~\cite{wang2022total}. The present framework offers a practical route to developing estimators for such differing physiological descriptions. Specifically, the one-time adaptation cost of 3~h per model on a single GPU compares with the 3~days required to train each tissue-specific CM from scratch, while preserving inference that is over two orders of magnitude faster than voxel-wise NNLS and five orders of magnitude faster than MCMC.
A limitation of the current framework is that the optimal LoRA rank and fine-tuning dataset size must be selected via grid search for each task, although our empirical experiments suggest that a rank of 4 and 100,000 training pairs are generally sufficient. Future work will focus on more extensive clinical validation across broader patient cohorts, radiotracers, and institutions, as well as investigating LoRACM as a standardization and harmonization tool across multi-center total-body PET scanners and protocols.



\section{Conclusion}
This study demonstrates that a pretrained CM for PET kinetic inference can be efficiently adapted to new protocols and tracers using LoRA. Requiring only a fraction of the original training data and just a few hours of fine-tuning, LoRACM successfully generated accurate total-body parametric images and Bayesian posteriors across three distinct clinical scenarios. By overcoming computational demands of training generative diffusion models from scratch, this framework provides a practical and scalable solution for extending rapid Bayesian kinetic inference to diverse dynamic PET applications.

\section*{Acknowledgment}
This work was supported by the National Imaging Facility (an NCRIS capability) at the Australian National Total Body PET Facility, The University of Sydney, funded by the NSW Government, and by the University of Sydney Core Research Facilities (CRF) Total Body PET Pilot Study Seed Funding Scheme for the pilot FP-CIT data acquisition.
The UC Davis data used in this work were acquired with support from the National Institutes of Health, USA.

\bibliographystyle{IEEEtran}
\bibliography{bib}

\end{document}